# Non-volatile silicon photonic memory with more than 4-bit per cell capability

X. Li[1], N. Youngblood[1], C.D. Wright[2], W. Pernice[3], and H. Bhaskaran[1,*]
[1]Department of Materials, University of Oxford, Oxford, UK, [2]Department of Engineering, University of Exeter, UK,
[3]Department of Physics, University of Münster, Germany, *email: harish.bhaskaran@materials.ox.ac.uk

*Abstract*—We present the first demonstration of an integrated photonic phase-change memory using $Ge_2Sb_2Te_5$ on silicon-on-insulator and demonstrate reliable multilevel operation with a single programming pulse. We also compare our results on silicon with previous demonstrations on silicon nitride. Crucially, achieving this on silicon enables tighter integration of traditional electronics with photonic memories in future, making phase-change photonic memory a viable and integrable technology.

## I. INTRODUCTION

Recently fully integrated photonic memory devices have been demonstrated on silicon nitride [1]. This approach not only enables photonic data storage on-chip, but also shows multilevel storage, improved signal-to-noise ratios (SNR), and reduced switching energy over available optical storage technology. Previous work on silicon nitride waveguides demonstrated both multilevel storage (≤10 levels) and computation in a fully optical framework [2]–[4]. This was despite the fact that silicon nitride does not benefit from a high refractive index contrast (~2/1.5 in $Si_3N_4/SiO_2$ vs 3.5/1.5 in $Si/SiO_2$) or the ability to integrate active photonics (such as modulators and photodetectors) directly in the waveguide. Silicon-On-Insulator (SOI), on the other hand, has both these advantages, leading to smaller footprint devices and easy integration with high-speed, CMOS-based electronics [5], [6]. Here, we demonstrate for the first time that it is possible to have non-volatile, phase-change optical memory on Si waveguides, which can significantly improve the operation speed of these devices compared with their $Si_3N_4$ counterparts. We also demonstrate a more efficient way of multilevel programming using double-step pulses for both WRITE and ERASE operations.

## II. DESIGN AND FABRICATION

To fabricate the waveguides, we use an SOI substrate with a 220 nm device layer and 3 µm buried oxide. Our device is illustrated in Fig. 1 where the phase-change material ($Ge_2Sb_2Te_5$ or "GST") is evanescently coupled to the optical mode of the Si waveguide, enabling all-optical WRITE, READ and ERASE operations. Electron beam lithography is used to pattern the device and dry etching is used for partially etch down Si waveguides (120 nm etch depth). Subsequently, a 10 nm functional layer of GST and a 10 nm protective layer of Indium-Tin-Oxide (ITO) are deposited on the top of the waveguide using RF sputtering (30 W RF power, 10 mTorr Ar atmosphere, and $2\times10^{-6}$ Torr base pressure). Fig. 2 shows optical and SEM images of the completed photonic memory cell. Light is coupled into the waveguide from optical fibers via grating couplers (shown in Fig. 2b).

Fig 3 shows the transmission of a bare silicon waveguide (blue curve), a silicon waveguide with a 4-µm-long amorphous GST cell (red curve), and the same cell after the GST has been crystallized (yellow curve). After crystallization, the loss increases significantly due to absorption by the GST which decreases the total transmission of the device. This change in transmission agrees well with our simulations of the GST/waveguide hybrid mode using a numerical eigenmode solver (Lumerical MODE Solver). Because of the higher refractive index, silicon waveguides provide better optical confinement of light than silicon nitride and therefore a smaller device footprint (compare Fig. 4 and 5). According to the calculated loss, for a 4-µm-long device, we expect to see the transmission decrease from 95% in the amorphous state (0.059 dB/µm) to 26% in the crystalline state (1.445 dB/µm). This is very similar to our experimental results in Fig. 3 where we see a decrease from 96% to 23% transmission after accounting for the transmission of the grating couplers.

## III. RESULTS AND DISCUSSION

### A. Comparison of Thermal Response

In order to understand the difference between the optical response of GST on silicon versus silicon nitride waveguides, we performed thermal and optical multi-physics simulations using FDTD and FEM analysis. Fig. 6 shows a top-down view of the simulated device temperature after a 20 ns optical pulse (200 pJ of absorbed energy) for both silicon and silicon nitride waveguides. In Fig. 7, we plot the time-dependent temperature rise for the area with the maximum temperature (denoted by a black dot in Fig. 6) and observe that while the GST on the silicon waveguide experiences a greater temperature rise initially, the peak temperature after the 20 ns pulse is lower than the GST on the silicon nitride waveguide. We attribute the higher temperature rise of the silicon device to the smaller device area and the lower final temperature to the significantly higher thermal conductivity of silicon compared to silicon nitride. This is also apparent in the very high cooling rating of the silicon device compared to the device on silicon nitride that occurs after 20 ns in Fig. 7.

We can directly measure the thermal response of our devices experimentally by monitoring the transmission of a low-power optical probe while sending WRITE pulses to the device. Because GST has a strong thermo-optic response [7], the transmission of the probe signal decreases with increasing temperature. Fig. 8 shows such a response in our silicon device where increasing the pulse length leads to a longer "dead time" in our device (defined as the time required to reach 1/e of the final transmission state). We plot the dead time as a function of

WRITE pulse duration in Fig. 9 and compare with previous measurements on silicon nitride. From these results, we can conclude that in order to efficiently switch the device on a silicon waveguide, for a given amount of WRITE energy, it is more beneficial to have a higher peak power and shorter WRITE time to prevent energy losses through heat diffusion. Using picosecond [8] or even femtosecond pulses [9] could therefore lead to the greatest efficiency gains in phase-change silicon photonics.

### B. Multilevel Photonic Memory on Silicon

In Fig. 10 we show non-volatile, multilevel operation of our devices where we vary the optical power of the WRITE pulse (25 ns duration) to achieve 29 distinct transmission levels. This is a significant improvement over previous multilevel demonstrations on silicon nitride waveguides [1]–[3]. The dependence on transmission contrast as a function of switching energy is shown in Fig. 11. We see that the energy required to achieve a similar transmission contrast in silicon compared to previous demonstrations [1] is greater by a factor of approximately 25. We attribute this to the higher thermal conductivity of our silicon waveguides (30 W/(m·K) in silicon nitride versus 143 W/(m·K) in silicon) and decreased evanescent coupling due to a higher mode confinement. The former can be overcome by reducing the duration of the WRITE pulse to less than a nanosecond, while the coupling can be enhanced through optimizing the device geometry.

By cycling the device multiple times with the same programming sequence, we plot the error between the expected transmission level and the actual level measured using an optical READ signal. The resulting histogram of the device is shown in Fig. 12 with a fitted Gaussian distribution. From the fit, we can see that the standard deviation of the error for all levels is 0.69% relative transmission.

### C. Single Pulse Recrystallization

In order to improve the efficiency of our devices, we investigated a single pulse method for fully recrystallizing the GST cell from any arbitrary amorphous state (see Fig. 13 for an illustration of this technique). Here we use a double-step ERASE pulse with an initial short, high power pulse followed by a longer, low power tail. This allows us to quickly heat the GST and hold it at the crystallization temperature until it reaches a fully crystalline state. Fig. 14 shows the thermo-optical response of our device during such pulses with increasing ERASE times. As the ERASE time increases, we enable the GST crystalline growth to occur for longer and to reach a greater degree of crystallinity. This can be seen in Fig. 15 and 16 where we plot the final state of the device as a function of ERASE time and the programming error associated with this method.

### D. Single Pulse Programming of Multilevel Storage

In a final demonstration, we use our double-step ERASE pulse to reach arbitrary transmission states regardless of the previous state of the material (Fig. 17). This reduces complexity of pulse sequencing for multilevel operation in photonic phase-change memory where pulses of decreasing energy are used to incrementally recrystallize small fractions of the GST cell until the desired level is reached (as in [1]). In the case of switching from a level of maximum transmission (i.e. amorphous) to a level of minimum transmission (i.e. fully crystalline), 19 pulses with energies ranging from 370 to 600 pJ were required corresponding to a total programming energy of 9.5 nJ and time of 3.8 µs. Similar approaches have also been demonstrated with picosecond and femtosecond pulses [2], [8], [9] but also require multiple recrystallization pulses of decreasing amplitude to reach the desired memory level. Here, we show multilevel operation using a single pulse ranging from 20 ns to 620 ns to achieve 30 distinct transmission levels without any knowledge of the prior state.

## IV. CONCLUSION

We have demonstrated silicon photonic phase-change memory using a single pulse programming method. This decreased the device footprint compared with silicon nitride photonics and reduced both the energy and time required for reaching arbitrary memory levels. We have characterized the reliability of this approach and outlined a comparison with other state-of-the-art programming methods. Our results are a significant step toward viable and integrable photonic phase-change memory devices on-chip.

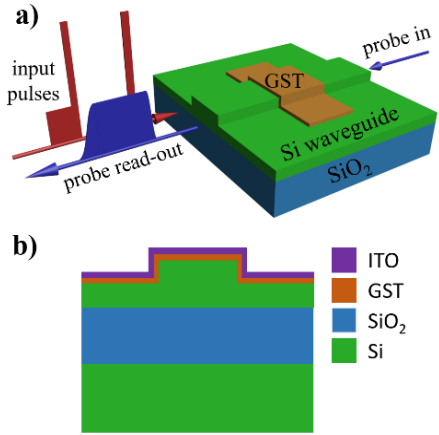

**Fig. 1.** (a) 3D illustration of integrated PCM photonic memory device and measurement scheme. Programming pulses are used to change the phase of the GST while a low power probe reads the transmission state. (b) Diagram of PCM photonic memory cell cross section showing GST on a shallow etched waveguide and ITO capping layer.

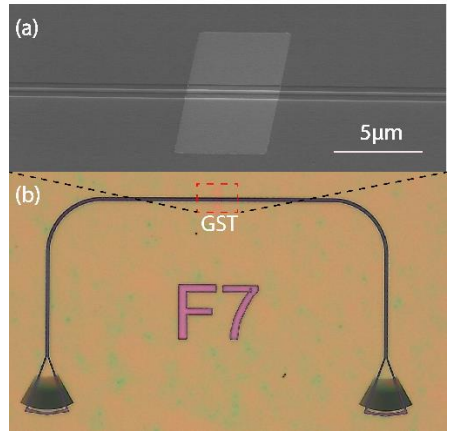

**Fig. 2.** (a) SEM close-up image of GST memory cell overlapping a silicon waveguide. (b) Optical microscope image of photonic circuit with input and output grating couplers and GST memory cell.

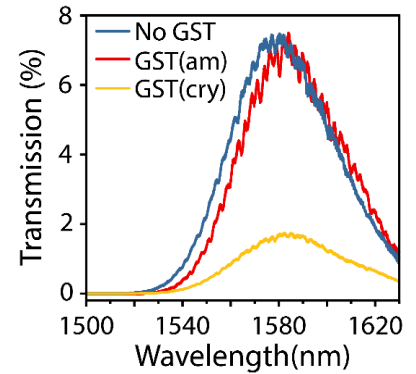

**Fig. 3.** Measured device transmission after the output grating coupler for a device before depositing GST (blue), with amorphous GST (red), and after crystallizing GST (yellow). The transmission significantly decreases after the GST crystallizes.

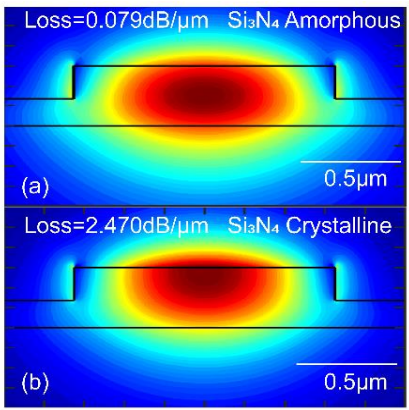

**Fig. 4.** Eigenmode simulation of fundamental TE optical mode propagating through the region with GST in the (a) amorphous and (b) crystalline states on a silicon nitride waveguide.

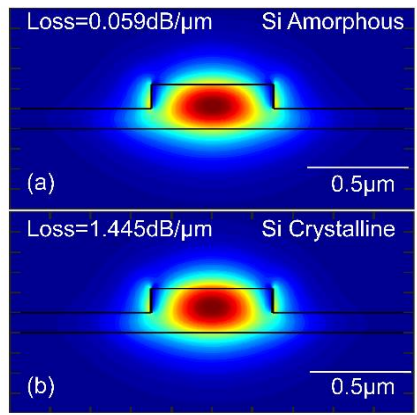

**Fig. 5.** Simulation of fundamental TE optical mode of a silicon waveguide with GST in the (a) amorphous and (b) crystalline state. The higher refractive index allows for a much smaller waveguide footprint compared to silicon nitride.

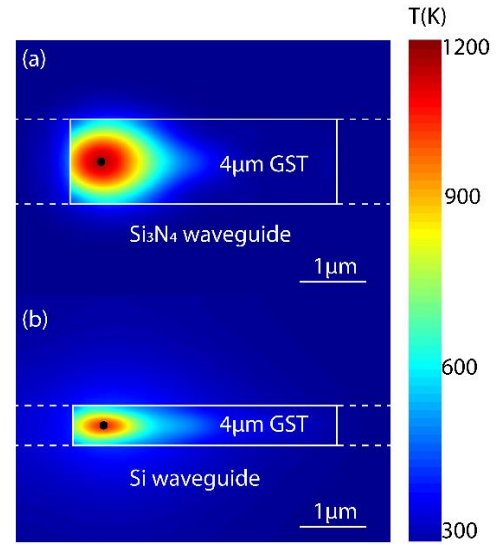

**Fig. 6.** Thermal simulations of a GST memory cell on (a) silicon nitride and (b) silicon waveguides after a 20 ns optical pulse (total absorbed energy of 200 pJ). The difference in the thermal conductivity of silicon nitride and silicon lead to different peak temperatures and distribution of heat throughout the GST.

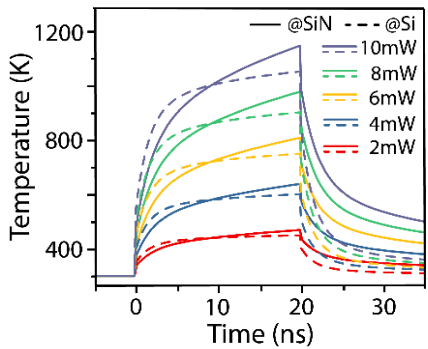

**Fig. 7.** Simulated peak temperature rise for GST on silicon nitride (solid lines) and silicon (dashed lines) waveguides at location of black dot in Fig. 6. Due to the smaller size of the silicon device and higher thermal conductivity, the temperature rises and falls more quickly compared to silicon nitride.

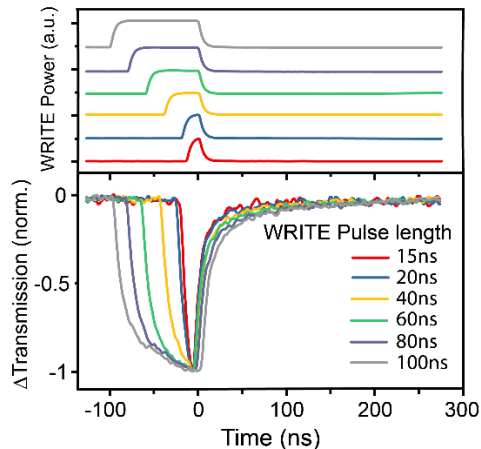

**Fig. 8.** Thermo-optic effect measured during and after a WRITE pulse. The initial dip during the WRITE pulse is due to the thermo-optic effect in GST which has higher absorption at higher temperatures. The remaining heat diffusing into the substrate also can be seen in the thermo-optic effect for approximately 100 ns after the WRITE pulse.

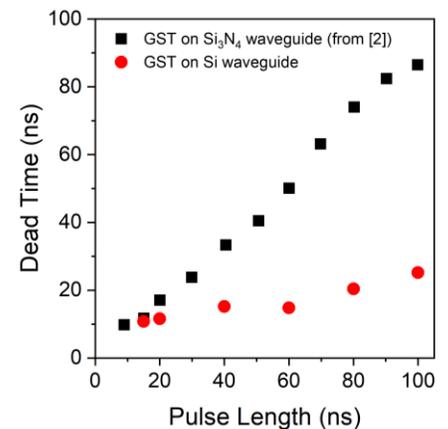

**Fig. 9.** Dead time in device (time required for the transmission to recover to 1/e of its final state) extracted from Fig. 8 and compared with data from ref. [2]. The high thermal conductivity of silicon significantly reduces dead time for long WRITE pulse lengths.

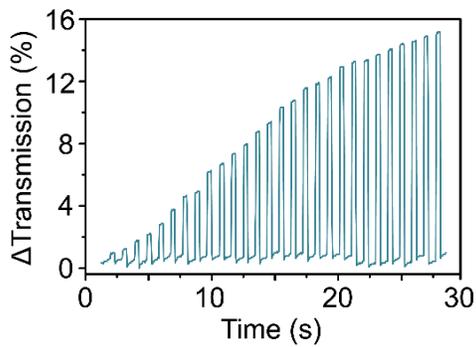

**Fig. 10.** Real time plot of the probe transmission versus time showing multilevel operation of photonic phase-change memory devices. 29 distinct and non-volatile levels can be achieved by controlling the power of the WRITE pulse.

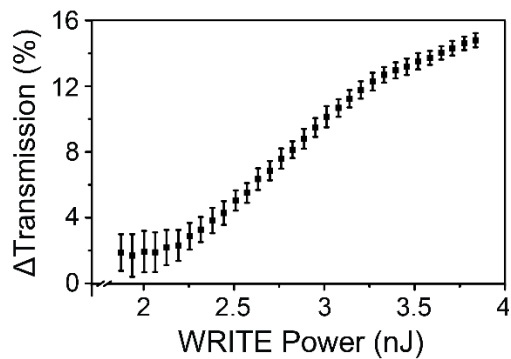

**Fig. 11.** Transmission level as a function of programming energy for a fixed 25 ns WRITE pulse. As the pulse energy increases, a larger region of crystalline GST is switched to the amorphous state which decreases absorption and increases transmission.

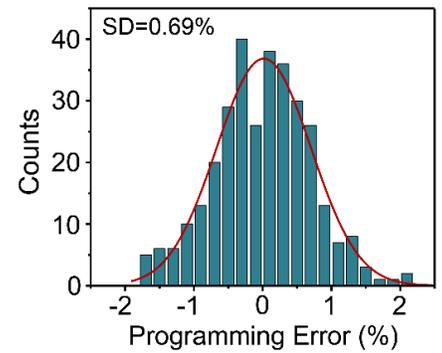

**Fig. 12.** Histogram of the programming error between the expected transmission level and the one which is experimentally measured. Multiple cycles were used to accumulate the error distribution. A Gaussian fit (red line) is used to calculate the standard deviation of the error.

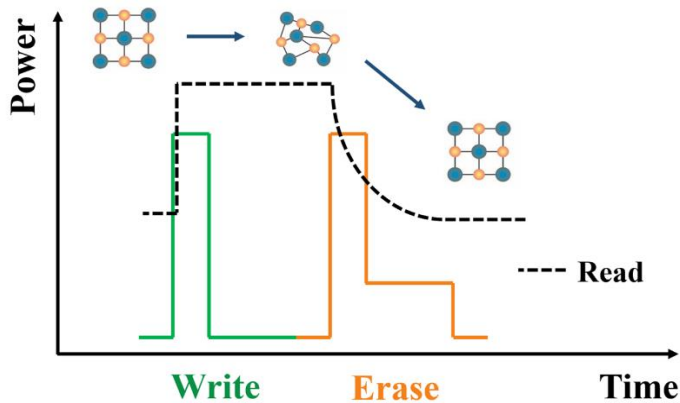

**Fig. 13.** Method for single pulse recrystallization. A double-step ERASE pulse is used to first bring the GST to a high temperature, then maintain the crystallization temperature while the material undergoes crystal growth.

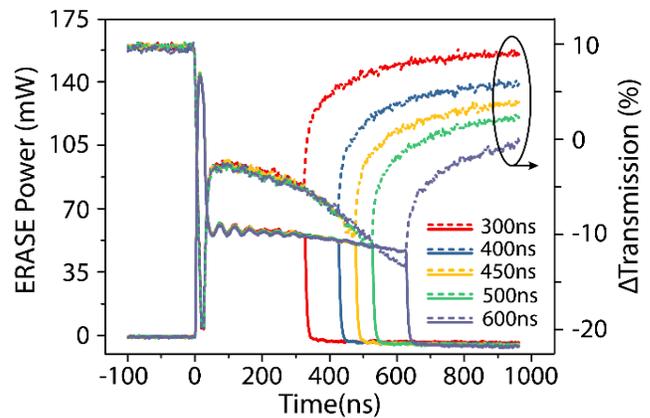

**Fig. 14.** Thermo-optic measurements of the GST transmission during a double-step ERASE pulse with increasing ERASE times. As the ERASE time increases from 300 to 600 ns, the GST reaches a state of increasing crystallization.

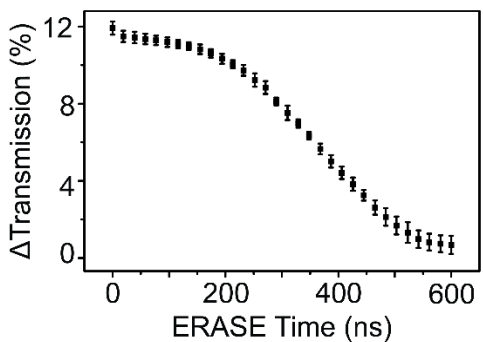

**Fig. 15.** Final transmission state as a function of ERASE time. Increasing the ERASE time leads to a lower final transmission state due to the time required for crystal growth to occur.

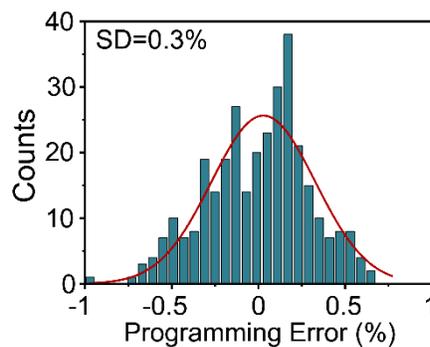

**Fig. 16.** Histogram of the programming error between the expected transmission level and the one which is experimentally measured by varying the ERASE time. Multiple cycles were used to accumulate the error distribution. A Gaussian fit (red line) is used to calculate the standard deviation of the error.

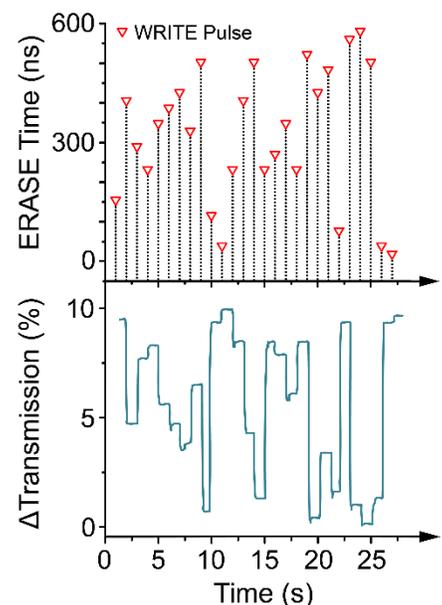

**Fig. 17.** Demonstration of single pulse, arbitrary level programming using a double-step ERASE pulse. The final transmission state of the device is independent of the previous state of the material.